\documentclass[aps,prl,twocolumn,groupedaddress,superscriptaddress,showpacs]{revtex4-1}

\usepackage{amsmath}
\usepackage{amssymb}
\usepackage{graphicx}
\usepackage{float}
\usepackage{color}

\begin{document}

\title{Interaction anisotropy and the KPZ to KPZQ transition in particle deposition at the edges of drying drops}

  \author{C. S. Dias}
   \email{csdias@fc.ul.pt}
    \affiliation{Departamento de F\'{\i}sica, Faculdade de Ci\^{e}ncias, Universidade de Lisboa, 
    1749-016 Lisboa, Portugal}
    \affiliation{Centro de F\'{i}sica Te\'{o}rica e Computacional, Universidade de Lisboa, 
    1749-016 Lisboa, Portugal}

  \author{P. J. Yunker}
   \email{peter.yunker@physics.gatech.edu}
    \affiliation{School of Physics, Georgia Institute of Technology, Atlanta, GA 30332, USA}

  \author{A. G. Yodh}
   \email{yodh@physics.upenn.edu}
    \affiliation{Department of Physics and Astronomy, University of Pennsylvania Philadelphia, PA 19104,USA}
    
  \author{N. A. M. Ara\'ujo}
   \email{nmaraujo@fc.ul.pt}
    \affiliation{Departamento de F\'{\i}sica, Faculdade de Ci\^{e}ncias, Universidade de Lisboa, 
    1749-016 Lisboa, Portugal}
    \affiliation{Centro de F\'{i}sica Te\'{o}rica e Computacional, Universidade de Lisboa, 
    1749-016 Lisboa, Portugal}
    
  \author{M. M. Telo da Gama}
   \email{mmgama@fc.ul.pt}
    \affiliation{Departamento de F\'{\i}sica, Faculdade de Ci\^{e}ncias, Universidade de Lisboa, 
    1749-016 Lisboa, Portugal}
    \affiliation{Centro de F\'{i}sica Te\'{o}rica e Computacional, Universidade de Lisboa, 
    1749-016 Lisboa, Portugal}

\begin{abstract}

The deposition process at the edge of evaporating colloidal drops varies with the shape of suspended particles. Experiments with prolate ellipsoidal particles suggest that the spatiotemporal properties of the deposit depend strongly on particle aspect ratio. As the aspect ratio increases, the particles form less densely-packed deposits and the statistical behavior of the deposit interface crosses over from the Kardar-Parisi-Zhang (KPZ) universality class to another universality class which was suggested to be consistent with the KPZ plus quenched disorder. Here, we numerically study the effect of particle interaction anisotropy on deposit growth. In essence, we model the ellipsoids, at the interface, as disk-like particles with two types of interaction patches that correspond to specific features at the poles and equator of the ellipsoid. The numerical results corroborate experimental observations and further suggest that the deposition transition can stem from interparticle interaction anisotropy. Possible extensions of our model to other systems are also discussed.
\end{abstract}

\maketitle

The ring-shaped deposit left by an evaporating colloidal drop, often called the coffee-ring effect, exhibits a rich phenomenology that has attracted attention for two decades 
\cite{Crivoi2015,Ma2011,Yang2014,Yunker2011,Yunker2013b,Deegan1997,Hu2002,Deegan2000a,Deegan2000,Marin2011,Weon2010,Kim2016a}, in part because it can affect the quality of coatings in the ink and printing industries \cite{Yang2014}. 
Briefly, the edge of the drying drop typically becomes pinned, producing radial fluid flows outward from the drop center to the drop edge. These convective flows, in turn, drag suspended particles to the drop edge where fascinating spatiotemporal deposition dynamics of particles occurs. 

\begin{figure}[t]
   \begin{center}
    \includegraphics[width=0.9\columnwidth]{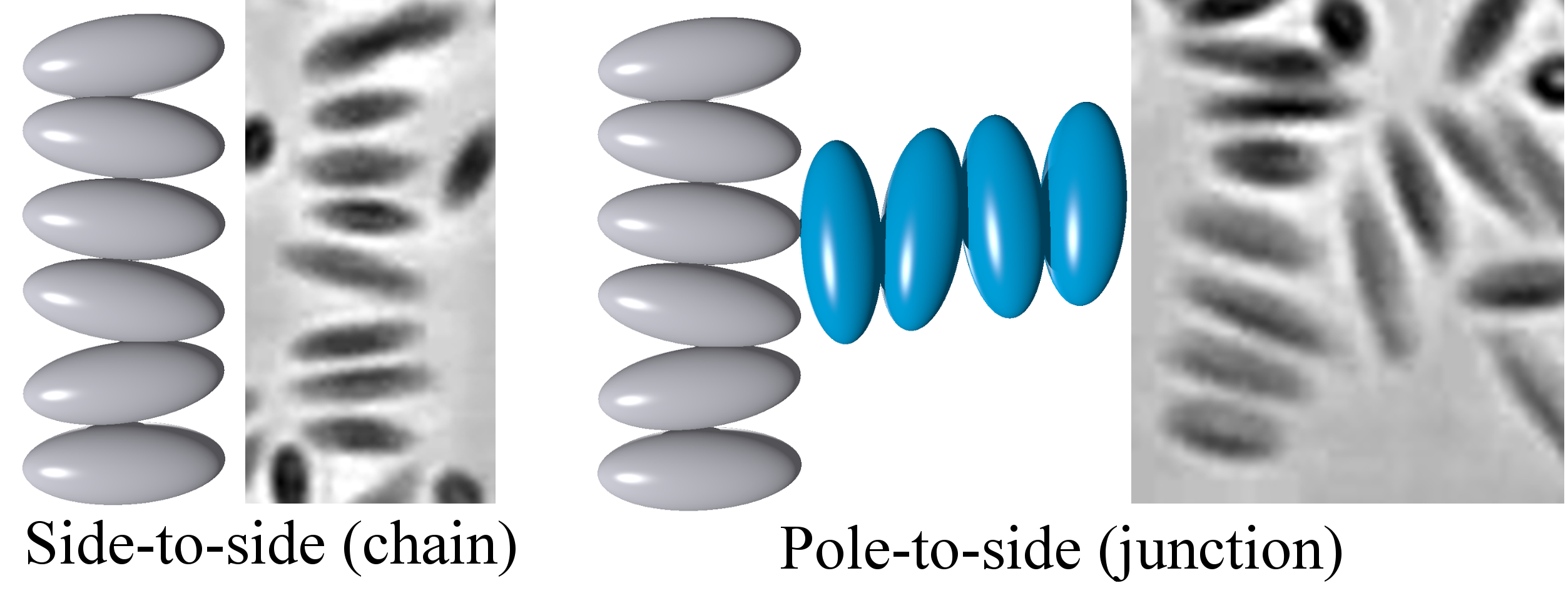} \\
\caption{Experiments with ellipsoidal (prolate) particles form two common structures:  chains and junctions.  The chains are characterized by many ellipsoidal particles arranged side- by-side. The junctions connect chains with different ellipsoid orientations, thereby forming new branches. Here, experimental snapshots are of ellipsoidal particles of aspect ratio
3.5 adsorbed at the air-water interface near the pinned edge of a drying one microliter drop of water.}
  \label{fig.figure1}
   \end{center}
  \end{figure}

\begin{figure}[t]
   \begin{center}
    \includegraphics[width=0.9\columnwidth]{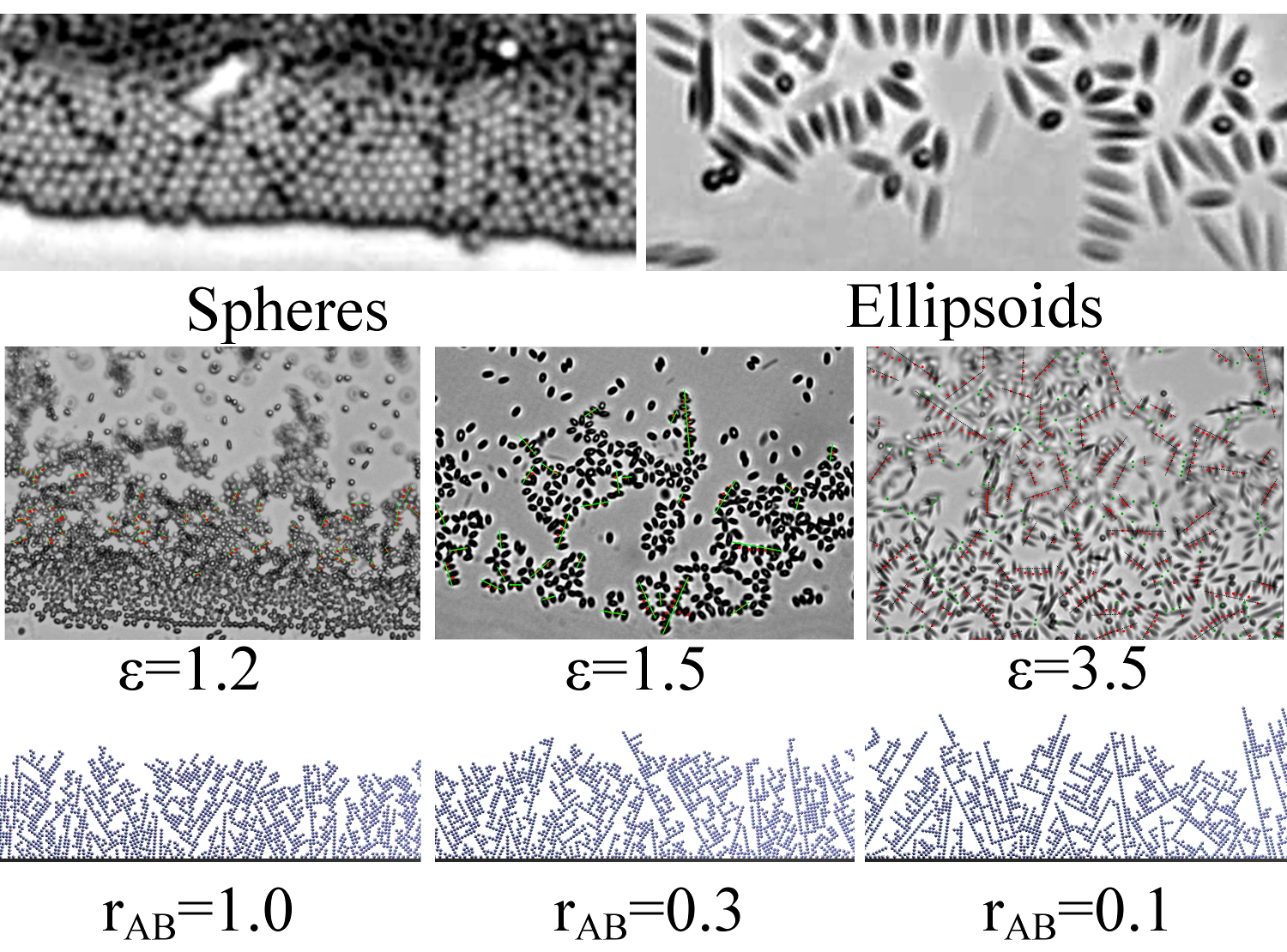} \\
\caption{Top: Experimental images of the drop edge for two limiting cases of the drying suspension, one with spherical and one with ellipsoidal (prolate) particles \cite{Yunker2013}. Middle: Experimental images of loosely packed networks of ellipsoidal polystyrene particles with aspect ratios 
$\varepsilon=\{1.2, 1.5, 3.5\}$; all the ellipsoids were stretched from the same colloidal spheres of diameter 1.3$\mu m$.
Bottom: Example snapshots of the deposition of $2A2B$ patchy-particles (from model simulations) obtained numerically, for ratios $r_{AB}=\{1.0, 0.3, 0.1\}$
(details in the text).
}
  \label{fig.figure2}
   \end{center}
  \end{figure}

Experiments have shown that the deposit morphology can depend strongly on suspended particle shape \cite{Yunker2013}. Sphere-like particles tend to pile up in a compact way, but highly anisotropic particles often form a loosely packed network consisting of chains and branches (see Fig.~\ref{fig.figure1}). Detailed analysis of the time evolution of these deposits revealed scaling properties of the growing interface that can strongly vary with particle aspect ratio, and that the structures are often kinetically trapped  \cite{Yunker2011, Yunker2013}. The statistics of the interfaces based on spherical particles is consistent with Poisson-like growth process for which lateral correlations decay exponentially with the distance. On the other hand, the statistics of interfaces based on anisotropic particles leads to self-affine roughness profiles (top row, Fig.~\ref{fig.figure2}).
Experimentally, the universality class of the latter interfacial roughening appears consistent with either Kardar-Parisi-Zhang (KPZ) \cite{Kardar1986} or Kardar-Parisi-Zhang with quenched disorder (KPZQ) \cite{Amaral1994}, depending on particle aspect ratio \cite{Yunker2013}. 

Although the initial observation of a KPZ process at the edges of drying colloidal drops was unexpected, theoretical studies have long suggested that the KPZ process is a robust and ubiquitous universality class for interfacial roughening. A few experiments have revealed signatures of KPZ processes (e.g., \cite{Wakita1997,Myllys1997a,Maunuksela1997}), and more recently, a comprehensive experimental observation of KPZ behavior was reported in a liquid crystal system \cite{Takeuchi2011,Takeuchi2010,DeNardis2017,Fukai2017}. By contrast to KPZ, the experimental suggestion of KPZQ behavior for a range of particle aspect ratios in the colloidal experiments was surprising. The KPZQ universality class is only expected to be observed at the critical point. Thus, the observations prompted questions about the mechanisms responsible, and in particular, about the crossover between the universality classes. Despite a recent effort along these lines \cite{Nicoli2013,Yunker2013c,Oliveira2014}, identification of the mechanism remains elusive. 

Here we employ simulations to investigate the effect of particle interaction potential anisotropy on spatiotemporal deposit growth processes. We model the ellipsoids, at the interface, as disk-like particles with two types of patches that correspond to anisotropic van der Waals interactions at the poles and equator of the ellipsoid. The numerical results corroborate experimental observations of a KPZ-to-KPZQ transition and suggest that the transition generally stems from interparticle interaction anisotropy.

\begin{figure}[t]
   \begin{center}
    \includegraphics[width=0.9\columnwidth]{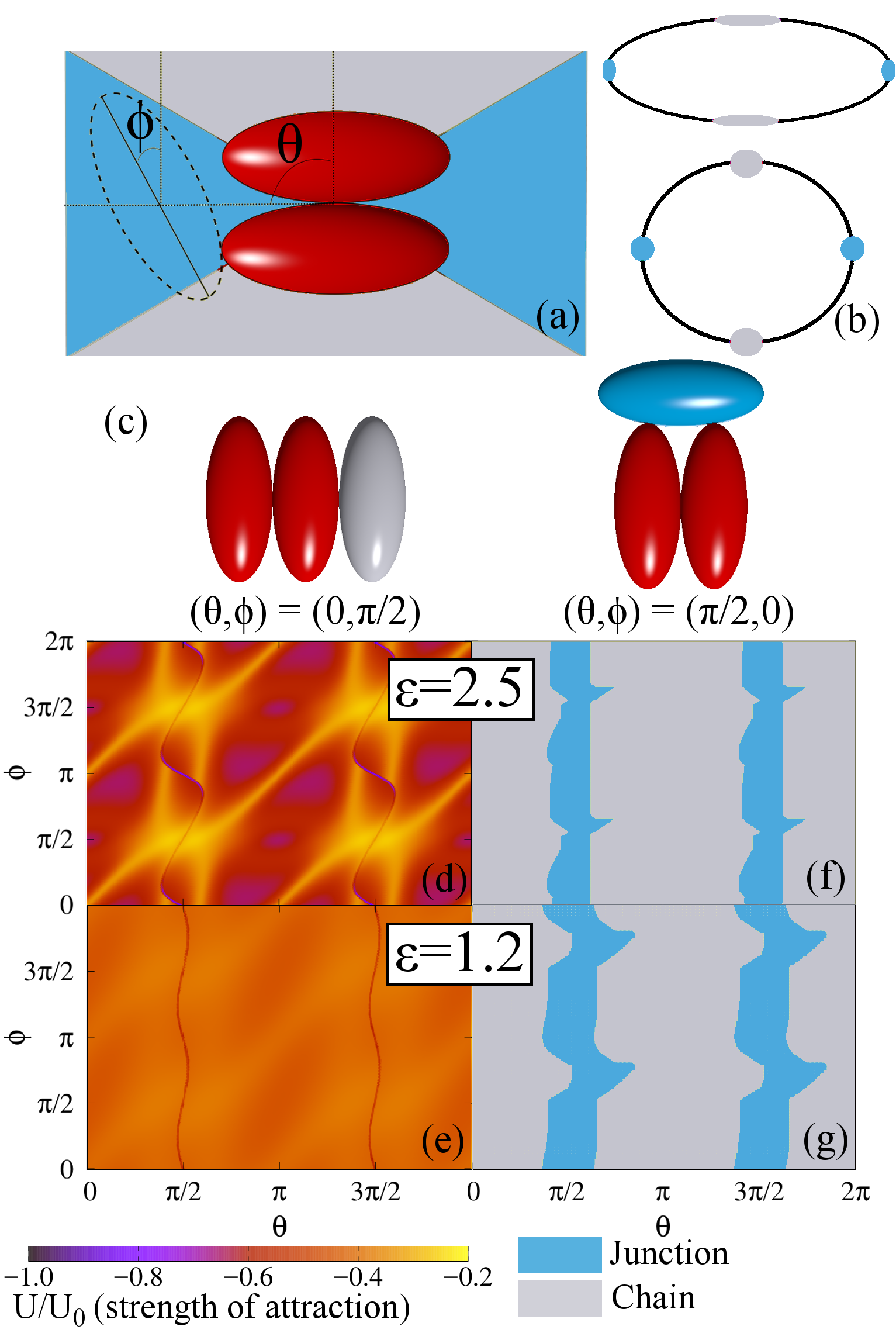} \\
\caption{(a) Scheme of the position of an ellipsoidal particle, relative to a pair of two other particles, 
uniquely identified by two angles: $\theta$ and $\phi$. The regions of different colors are 
the attraction basins of the chain (light-gray) and junction (light-blue) configurations. 
(b) Schematic representation of the mapping of the ellipsoid (top)
on a 2A2B patchy particle (bottom). 
(c) Schematic representation of the configurations corresponding to the minima of the energy landscape, namely, 
of chain and junction configurations.
(d,e) Energy landscapes calculated from the potential given by Eq.~(\ref{eq.Derjaguin}), defined on the two-parameter space, $\theta$ and $\phi$,
for aspect ratios of $\varepsilon=2.5$ and $\varepsilon=1.2$, respectively, the scale is in units of $U/U_0$. 
(f,g) attraction basins for the chain and junction configurations (each attraction basin is colored as in (a) gray for three-particle chains and blue for three-particle junctions), for aspect ratios $\varepsilon=2.5$ and $\varepsilon=1.2$, respectively.} 
  \label{fig.figure3}
   \end{center}
  \end{figure}

In the colloidal drop experiments, the strong anisotropic interparticle interactions play a critical role. To simulate these anisotropic interactions, at the interface, we model the ellipsoids as disk-like particles with four interaction patches: two A-patches (red) and two B-patches (green), corresponding to the sides and the poles of the ellipsoid, respectively (see Fig.~\ref{fig.figure3}(b)). 
The simulated particles are transported ballistically towards a substrate, one at a time. They form bonds with other particles solely through the contacting patches. Three binding probabilities are considered, depending on the pair of interacting patches, namely, $P_{AA}$ (for $A-A$), $P_{AB}$ (for $A-B$), and $P_{BB}$ (for $B-B$). Note that $P_{AA}$ and $P_{AB}$ are related to the probability of forming a chain and a junction, respectively (see Fig.~\ref{fig.figure3}(c)). Without loss of generality, a sticking coefficient $r_{AB} = P_{AB}/P_{AA}$, can also be defined. To access the relevant timescales, we refrain from carrying out detailed molecular dynamics simulations and propose a stochastic growth model instead.

The pairwise interactions in the experiments are very strong ($\approx 10^5 k_BT$) \cite{Botto2012,Davies2014,Loudet2005}, and it is rare that two touching particles will ever separate. Thus, in our model we assume that the local relaxation is driven by the minimization of the pairwise interaction energy between the particles in a steepest descent towards the local minimum. In this case, the two particles will remain in con- tact at all times while they slide past one another to find energy minima. Given the quadrupolar symmetry of the interaction, the local minima of the pairwise interaction corresponds either to a chain (side-to-side) or a junction (pole-to-side) configuration, as schematically represented in Figs.~\ref{fig.figure1}~and~\ref{fig.figure3}(c). The probability that a pair of particles will contact and relax towards one of those two configurations is given by the relative size of the corresponding basins of attraction.

To delineate the basins of attraction and calculate their relative sizes, we computed the energy landscape of two such colloidal particles at contact. At short distances, relevant interactions are readily modeled by van der Waals-like interactions \cite{Lehle2008} within the Derjaguin (or proximity) approximation, which can be written as, 
\begin{equation}
 U=-U_0\sqrt{\frac{2}{C_1+C_2}}, \label{eq.Derjaguin}
\end{equation}
where $C_1$ and $C_2$ are the local curvatures of the two ellipsoids at the point of contact and $U_0$ is an energy parameter that depends on the contact distance between the surface of the particles, which we assume constant. For simplicity, we have considered only configurations where the major axis of all ellipsoids is in the same plane. The contact points between two ellipsoids and the corresponding local curvatures are computed using the Perram and Wertheim method
\cite{Perram1985}. 

We discretized the two-parameter ($\theta$,$\phi$) space into a mesh of possible contact
points between an ellipsoid and a pair of ellipsoids, as schematically shown in Fig.~\ref{fig.figure3}(a), and we compute the entire energy landscape, as shown in  Figs.~\ref{fig.figure3}(d) and (e), for two different aspect ratios. A basin includes all possible initial configurations that lead to the local minimum via a steepest descent, with constraint that ellipsoids are always in contact. This discretized energy landscape is a ranked surface \cite{Araujo2015b}, 
and the size of each attraction basin is obtained by using the flooding algorithm proposed in Ref.~\cite{Araujo2015b}. 
Figures~\ref{fig.figure3}(f) and (g) show the corresponding attraction basins for the chain (light) and junction (dark) minima derived from ellipsoids with two different aspect ratios: $\varepsilon=2.5$ and $\varepsilon=1.2$. Notice that the larger aspect ratios lead to narrower basins of attraction for the junction configuration; this suggests a lower probability for forming junctions compared to forming chains. This finding is consistent with the observed increasing chain size with $\varepsilon$ \cite{Yunker2013}.

To parametrize the sticking coefficient, $r_{AB}$, we consider the size of the basins associated with each of the local energy minima (chain and junction). 
The $r_{AB}(\varepsilon)$, for a certain aspect ratio $\varepsilon$, is defined as,
\begin{equation}
 r_{AB}=A_\mathrm{junction}/A_\mathrm{chain}, \label{eq.rab}
\end{equation}
where $A_\mathrm{junction}$ and $A_\mathrm{chain}$ are the areas of the attraction basins corresponding to the junction (darker color) and chain (lighter color)
local minima, respectively. Since, the pole-to-side, ($\theta$, $\phi$)=($\pi/2$,0),
and pole-to-pole, ($\theta$, $\phi$)=($\pi/2$,$\pi/2$), configurations have attraction basins of similar size, we set $P_{BB}=P_{AB}$.
From Figs.~\ref{fig.figure3}(f) and (g), we see that $r_\mathrm{AB}$ is smaller for the smaller aspect ratio.

The simulations reveal that lower $r_\mathrm{AB}$ leads to more loosely packed structures (Fig.~\ref{fig.figure2}, bottom row), in qualitative agreement with the experiments.
Further, our model corresponds to the irreversible adsorption of two- type-patch colloids on substrates \cite{Dias2014a}, a system for which the very same transition in the kinetic roughening was reported in Ref.~\cite{Araujo2015}. For significantly dissimilar binding probabilities ($r_{AB}\ll 1$), the interface is in the KPZQ universality class,
while for $r_{AB}\approx 1$, it tumbles into the robust KPZ universality class. 

\begin{figure}[t]
   \begin{center}
    \includegraphics[width=0.9\columnwidth]{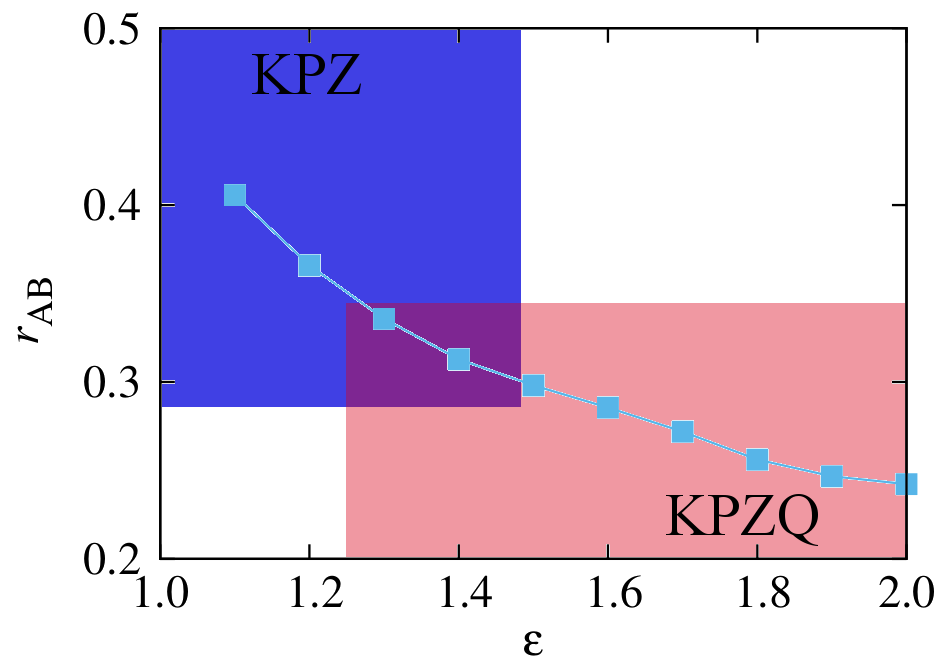} \\
\caption{Sticking coefficient $r_{AB}$ as a function of the particle aspect ratio $\varepsilon$. The colored regions indicate where the KPZ (blue) and KPZQ (red) are observed both experimentally and theoretically. The overlap region (purple) represents the crossover region between KPZ and KPZQ, both experimentally and numerically.}
  \label{fig.figure4}
   \end{center}
  \end{figure}

Figure~\ref{fig.figure4} shows the curve of $r_\mathrm{AB}$ as a function of $\varepsilon$. The colored regions correspond to the ones where KPZ (blue) and KPZQ (red) are identified both numerically and experimentally. The crossover region between KPZ and KPZQ, observed experimentally and numerically, is where the two colored regions overlap. Notably, the transition in the experimental system occurs for a similar range of parameters predicted by the theoretical model. In practice, finite-size effects and experimental error likely broaden the apparent crossover regime \cite{Dias2014a}. The particles utilized in the experiments are polydisperse in aspect ratio; this polydispersity would be expected to broaden the experimentally observed crossover as well \cite{Yunker2013}.

\begin{figure}[t]
    \begin{center}
     \includegraphics[width=0.9\columnwidth]{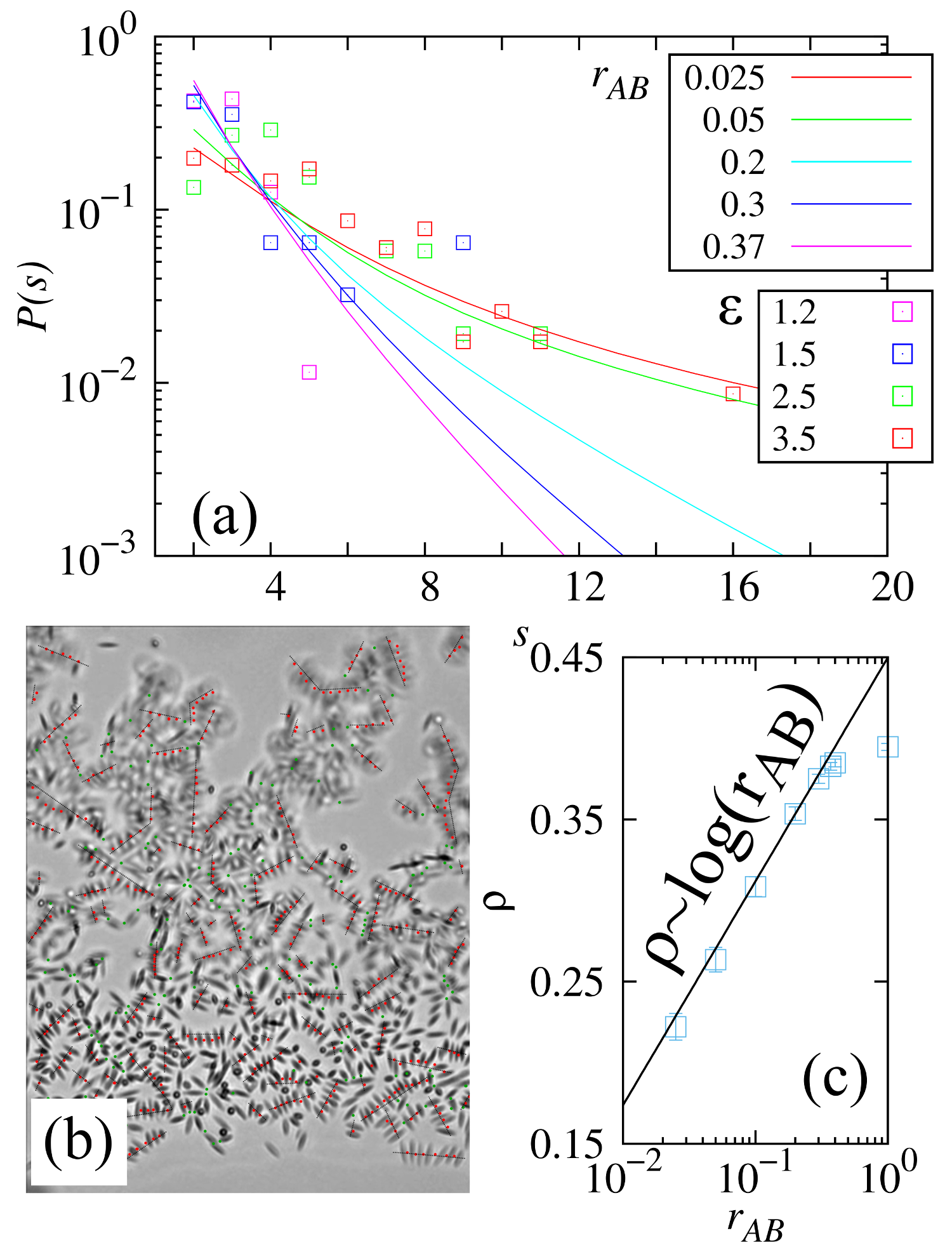} \\
 \caption{(a) Direct measurement of the distribution function $P(s)$ of 
 chain sizes $s$ of ellipsoidal colloids taken from experimental images at aspect ratios
 of $\varepsilon=\{1.2, 1.5, 2.5, 3.5\}$ and numerical results at the mapped $r_{AB}$ values, for the same color. 
 (inset) Snapshot of an experimental network of ellipsoidal colloids with an aspect ratio of $\varepsilon=3.5$.
 (b) Numerically obtained area fraction, $\frac{N\pi r^2}{Lh}$, as a function of $r_{AB}$, where $N$ is the number
 of particles, $r$ is the radius of the particle cross section, $L$ the lateral size, and $h$ the average height .}
   \label{fig.figure5}
    \end{center}
   \end{figure}

To facilitate further comparisons between the simulations and experiments, we also measured the distribution of chain sizes. Chains are defined as sequences of particles between consecutive junctions. Figure~\ref{fig.figure5}(a) shows the distribution of chain sizes, where the experimental values were obtained by directly counting the particles in the experimental images. Again, despite large experimental error bars, the numerical and experimental results are consistent. 

Figure~\ref{fig.figure5}(b) shows also the numerical area fraction (cross section of the particle over the observed area), 
$\frac{A_{particles}}{A_{observed}}=\frac{N\pi r^2}{Lh}$, as a function of $r_{AB}$, where $N$ is the number
of particles, $r$ is the radius of the particle cross section, $L$ the lateral substrate size, 
and $h$ the average height. For lower values of $r_{AB}$, the density scales logarithmically, well inside the KPZQ region. Experimental measurements show a trend that is in agreement with the model, albeit over a narrower region.  In the experiments, large scale rearrangements occasionally occur. Thus, we measured the area fraction in pristine regions, wherein no rearrangements occur after particles are deposited. The area fraction values computed from image analysis of the experiments are $\rho=\{0.532\pm0.038; 0.518\pm0.035; 0.478\pm0.030\}$ for aspect ratios of $\varepsilon=\{1.2; 1.5; 3.5\}$
respectively. While slightly higher than the model, several factors might justify the
quantitative differences. For example, numerically, particles are monodisperse disks. By contrast, in the experiments, particles are polydisperse ellipsoids. Thus, in the experiments, the distance between the centers of two touching particles depends on the local configuration, which will affect the packing fraction.

Using numerical simulations, we have identified an underlying mechanism responsible for the experimentally observed transition in interfacial roughening universality class from KPZ to KPZQ. In the colloidal experiments, the anisotropy arose because of shape-dependent interactions; in the numerical simulations, interaction patches with different (anisotropic) strengths are shown to drive the effect. In the present work, we established a relationship between the aspect ratio of the colloidal particles and the corresponding probability of forming side-to-side (chain) versus  pole-to-side (junction) configurations. The conditions for which the transition are predicted by the theoretical model is comparable to those observed experimentally. With this mapping, we provide a potential mechanistic justification for the difference in universality class based on directional anisotropy in interactions. In addition, we developed a simple theoretical tool that permits numerical access to time and length scales of interest in experiment. The methods developed here have potential applications beyond this system. In particular, the mapping of anisotropic particles onto spherical patch particles opens the possibility of new theoretical studies. For instance, it was shown recently that even for spherical particles, the particle roughness can lead to anisotropic interactions~\cite{Zanini2017}. It would be interesting to verify if the same kinds of transitions are observed as a function of the particle roughness. Also, different particle shapes will induce other symmetries in the anisotropy of the particle-particle interaction. In those cases, the mapping proposed here will need to include a different number of patches. In principle, binary mixtures of particles can be considered in a similar fashion. Thus, the framework discussed here may provide a route to identify the optimal particle shape for a desired deposit morphology, providing a novel degree of control over the density and structure of the final deposit. 

\begin{acknowledgments}
We acknowledge financial support from the
Portuguese Foundation for Science and Technology (FCT) under Contracts
nos. EXCL/FIS-NAN/0083/2012, UID/FIS/00618/2013, and IF/00255/2013. 
AGY acknowledges partial support from the NSF through the DMR-160738 and from NASA through NNX08AO0G. 
PJY acknowledges partial support from the Georgia Tech Soft Matter Incubator.

\end{acknowledgments}

\bibliography{paper}

\begin{thebibliography}{34}%
\makeatletter
\providecommand \@ifxundefined [1]{%
 \@ifx{#1\undefined}
}%
\providecommand \@ifnum [1]{%
 \ifnum #1\expandafter \@firstoftwo
 \else \expandafter \@secondoftwo
 \fi
}%
\providecommand \@ifx [1]{%
 \ifx #1\expandafter \@firstoftwo
 \else \expandafter \@secondoftwo
 \fi
}%
\providecommand \natexlab [1]{#1}%
\providecommand \enquote  [1]{``#1''}%
\providecommand \bibnamefont  [1]{#1}%
\providecommand \bibfnamefont [1]{#1}%
\providecommand \citenamefont [1]{#1}%
\providecommand \href@noop [0]{\@secondoftwo}%
\providecommand \href [0]{\begingroup \@sanitize@url \@href}%
\providecommand \@href[1]{\@@startlink{#1}\@@href}%
\providecommand \@@href[1]{\endgroup#1\@@endlink}%
\providecommand \@sanitize@url [0]{\catcode `\\12\catcode `\$12\catcode
  `\&12\catcode `\#12\catcode `\^12\catcode `\_12\catcode `\%12\relax}%
\providecommand \@@startlink[1]{}%
\providecommand \@@endlink[0]{}%
\providecommand \url  [0]{\begingroup\@sanitize@url \@url }%
\providecommand \@url [1]{\endgroup\@href {#1}{\urlprefix }}%
\providecommand \urlprefix  [0]{URL }%
\providecommand \Eprint [0]{\href }%
\providecommand \doibase [0]{http://dx.doi.org/}%
\providecommand \selectlanguage [0]{\@gobble}%
\providecommand \bibinfo  [0]{\@secondoftwo}%
\providecommand \bibfield  [0]{\@secondoftwo}%
\providecommand \translation [1]{[#1]}%
\providecommand \BibitemOpen [0]{}%
\providecommand \bibitemStop [0]{}%
\providecommand \bibitemNoStop [0]{.\EOS\space}%
\providecommand \EOS [0]{\spacefactor3000\relax}%
\providecommand \BibitemShut  [1]{\csname bibitem#1\endcsname}%
\let\auto@bib@innerbib\@empty
%</preamble>
\bibitem [{\citenamefont {Crivoi}\ \emph {et~al.}(2015)\citenamefont {Crivoi},
  \citenamefont {Zhong},\ and\ \citenamefont {Duan}}]{Crivoi2015}%
  \BibitemOpen
  \bibfield  {author} {\bibinfo {author} {\bibfnamefont {A.}~\bibnamefont
  {Crivoi}}, \bibinfo {author} {\bibfnamefont {X.}~\bibnamefont {Zhong}}, \
  and\ \bibinfo {author} {\bibfnamefont {F.}~\bibnamefont {Duan}},\ }\href@noop
  {} {\bibfield  {journal} {\bibinfo  {journal} {Phys. Rev. E}\ }\textbf
  {\bibinfo {volume} {92}},\ \bibinfo {pages} {032302} (\bibinfo {year}
  {2015})}\BibitemShut {NoStop}%
\bibitem [{\citenamefont {Ma}\ and\ \citenamefont {Hao}(2011)}]{Ma2011}%
  \BibitemOpen
  \bibfield  {author} {\bibinfo {author} {\bibfnamefont {H.}~\bibnamefont
  {Ma}}\ and\ \bibinfo {author} {\bibfnamefont {J.}~\bibnamefont {Hao}},\
  }\href@noop {} {\bibfield  {journal} {\bibinfo  {journal} {Chem. Soc. Rev.}\
  }\textbf {\bibinfo {volume} {40}},\ \bibinfo {pages} {5457} (\bibinfo {year}
  {2011})}\BibitemShut {NoStop}%
\bibitem [{\citenamefont {Yang}\ \emph {et~al.}(2014)\citenamefont {Yang},
  \citenamefont {Li},\ and\ \citenamefont {Sun}}]{Yang2014}%
  \BibitemOpen
  \bibfield  {author} {\bibinfo {author} {\bibfnamefont {X.}~\bibnamefont
  {Yang}}, \bibinfo {author} {\bibfnamefont {C.~Y.}\ \bibnamefont {Li}}, \ and\
  \bibinfo {author} {\bibfnamefont {Y.}~\bibnamefont {Sun}},\ }\href@noop {}
  {\bibfield  {journal} {\bibinfo  {journal} {Soft Matt.}\ }\textbf {\bibinfo
  {volume} {10}},\ \bibinfo {pages} {4458} (\bibinfo {year}
  {2014})}\BibitemShut {NoStop}%
\bibitem [{\citenamefont {Yunker}\ \emph {et~al.}(2011)\citenamefont {Yunker},
  \citenamefont {Still}, \citenamefont {Lohr},\ and\ \citenamefont
  {Yodh}}]{Yunker2011}%
  \BibitemOpen
  \bibfield  {author} {\bibinfo {author} {\bibfnamefont {P.~J.}\ \bibnamefont
  {Yunker}}, \bibinfo {author} {\bibfnamefont {T.}~\bibnamefont {Still}},
  \bibinfo {author} {\bibfnamefont {M.~A.}\ \bibnamefont {Lohr}}, \ and\
  \bibinfo {author} {\bibfnamefont {A.~G.}\ \bibnamefont {Yodh}},\ }\href@noop
  {} {\bibfield  {journal} {\bibinfo  {journal} {Nature}\ }\textbf {\bibinfo
  {volume} {476}},\ \bibinfo {pages} {308} (\bibinfo {year}
  {2011})}\BibitemShut {NoStop}%
\bibitem [{\citenamefont {Yunker}\ \emph
  {et~al.}(2013{\natexlab{a}})\citenamefont {Yunker}, \citenamefont {Durian},\
  and\ \citenamefont {Yodh}}]{Yunker2013b}%
  \BibitemOpen
  \bibfield  {author} {\bibinfo {author} {\bibfnamefont {P.~J.}\ \bibnamefont
  {Yunker}}, \bibinfo {author} {\bibfnamefont {D.~J.}\ \bibnamefont {Durian}},
  \ and\ \bibinfo {author} {\bibfnamefont {A.~G.}\ \bibnamefont {Yodh}},\
  }\href@noop {} {\bibfield  {journal} {\bibinfo  {journal} {Physics Today}\
  }\textbf {\bibinfo {volume} {66}},\ \bibinfo {pages} {60} (\bibinfo {year}
  {2013}{\natexlab{a}})}\BibitemShut {NoStop}%
\bibitem [{\citenamefont {Deegan}\ \emph {et~al.}(1997)\citenamefont {Deegan},
  \citenamefont {Bakajin}, \citenamefont {Dupont}, \citenamefont {Huber},
  \citenamefont {Nagel},\ and\ \citenamefont {Witten}}]{Deegan1997}%
  \BibitemOpen
  \bibfield  {author} {\bibinfo {author} {\bibfnamefont {R.~D.}\ \bibnamefont
  {Deegan}}, \bibinfo {author} {\bibfnamefont {O.}~\bibnamefont {Bakajin}},
  \bibinfo {author} {\bibfnamefont {T.~F.}\ \bibnamefont {Dupont}}, \bibinfo
  {author} {\bibfnamefont {G.}~\bibnamefont {Huber}}, \bibinfo {author}
  {\bibfnamefont {S.~R.}\ \bibnamefont {Nagel}}, \ and\ \bibinfo {author}
  {\bibfnamefont {T.~A.}\ \bibnamefont {Witten}},\ }\href@noop {} {\bibfield
  {journal} {\bibinfo  {journal} {Nature}\ }\textbf {\bibinfo {volume} {389}},\
  \bibinfo {pages} {827} (\bibinfo {year} {1997})}\BibitemShut {NoStop}%
\bibitem [{\citenamefont {Hu}\ and\ \citenamefont {Larson}(2002)}]{Hu2002}%
  \BibitemOpen
  \bibfield  {author} {\bibinfo {author} {\bibfnamefont {H.}~\bibnamefont
  {Hu}}\ and\ \bibinfo {author} {\bibfnamefont {R.~G.}\ \bibnamefont
  {Larson}},\ }\href@noop {} {\bibfield  {journal} {\bibinfo  {journal} {J
  Phys. Chem. B}\ }\textbf {\bibinfo {volume} {106}},\ \bibinfo {pages} {1334}
  (\bibinfo {year} {2002})}\BibitemShut {NoStop}%
\bibitem [{\citenamefont {Deegan}(2000)}]{Deegan2000a}%
  \BibitemOpen
  \bibfield  {author} {\bibinfo {author} {\bibfnamefont {R.~D.}\ \bibnamefont
  {Deegan}},\ }\href@noop {} {\bibfield  {journal} {\bibinfo  {journal} {Phys.
  Rev. E}\ }\textbf {\bibinfo {volume} {61}},\ \bibinfo {pages} {475} (\bibinfo
  {year} {2000})}\BibitemShut {NoStop}%
\bibitem [{\citenamefont {Deegan}\ \emph {et~al.}(2000)\citenamefont {Deegan},
  \citenamefont {Bakajin}, \citenamefont {Dupont}, \citenamefont {Huber},
  \citenamefont {Nagel},\ and\ \citenamefont {Witten}}]{Deegan2000}%
  \BibitemOpen
  \bibfield  {author} {\bibinfo {author} {\bibfnamefont {R.~D.}\ \bibnamefont
  {Deegan}}, \bibinfo {author} {\bibfnamefont {O.}~\bibnamefont {Bakajin}},
  \bibinfo {author} {\bibfnamefont {T.~F.}\ \bibnamefont {Dupont}}, \bibinfo
  {author} {\bibfnamefont {G.}~\bibnamefont {Huber}}, \bibinfo {author}
  {\bibfnamefont {S.~R.}\ \bibnamefont {Nagel}}, \ and\ \bibinfo {author}
  {\bibfnamefont {T.~A.}\ \bibnamefont {Witten}},\ }\href@noop {} {\bibfield
  {journal} {\bibinfo  {journal} {Phys. Rev. E}\ }\textbf {\bibinfo {volume}
  {62}},\ \bibinfo {pages} {756} (\bibinfo {year} {2000})}\BibitemShut
  {NoStop}%
\bibitem [{\citenamefont {Mar{\'{i}}n}\ \emph {et~al.}(2011)\citenamefont
  {Mar{\'{i}}n}, \citenamefont {Gelderblom}, \citenamefont {Lohse},\ and\
  \citenamefont {Snoeijer}}]{Marin2011}%
  \BibitemOpen
  \bibfield  {author} {\bibinfo {author} {\bibfnamefont {A.~G.}\ \bibnamefont
  {Mar{\'{i}}n}}, \bibinfo {author} {\bibfnamefont {H.}~\bibnamefont
  {Gelderblom}}, \bibinfo {author} {\bibfnamefont {D.}~\bibnamefont {Lohse}}, \
  and\ \bibinfo {author} {\bibfnamefont {J.~H.}\ \bibnamefont {Snoeijer}},\
  }\href@noop {} {\bibfield  {journal} {\bibinfo  {journal} {Phys. Rev. Lett.}\
  }\textbf {\bibinfo {volume} {107}},\ \bibinfo {pages} {085502} (\bibinfo
  {year} {2011})}\BibitemShut {NoStop}%
\bibitem [{\citenamefont {Weon}\ and\ \citenamefont {Je}(2010)}]{Weon2010}%
  \BibitemOpen
  \bibfield  {author} {\bibinfo {author} {\bibfnamefont {B.~M.}\ \bibnamefont
  {Weon}}\ and\ \bibinfo {author} {\bibfnamefont {J.~H.}\ \bibnamefont {Je}},\
  }\href@noop {} {\bibfield  {journal} {\bibinfo  {journal} {Phys. Rev. E}\
  }\textbf {\bibinfo {volume} {82}},\ \bibinfo {pages} {015305} (\bibinfo
  {year} {2010})}\BibitemShut {NoStop}%
\bibitem [{\citenamefont {Kim}\ \emph {et~al.}(2016)\citenamefont {Kim},
  \citenamefont {Boulogne}, \citenamefont {Um}, \citenamefont {Jacobi},
  \citenamefont {Button},\ and\ \citenamefont {Stone}}]{Kim2016a}%
  \BibitemOpen
  \bibfield  {author} {\bibinfo {author} {\bibfnamefont {H.}~\bibnamefont
  {Kim}}, \bibinfo {author} {\bibfnamefont {F.}~\bibnamefont {Boulogne}},
  \bibinfo {author} {\bibfnamefont {E.}~\bibnamefont {Um}}, \bibinfo {author}
  {\bibfnamefont {I.}~\bibnamefont {Jacobi}}, \bibinfo {author} {\bibfnamefont
  {E.}~\bibnamefont {Button}}, \ and\ \bibinfo {author} {\bibfnamefont {H.~A.}\
  \bibnamefont {Stone}},\ }\href@noop {} {\bibfield  {journal} {\bibinfo
  {journal} {Phys. Rev. Lett.}\ }\textbf {\bibinfo {volume} {116}},\ \bibinfo
  {pages} {124501} (\bibinfo {year} {2016})}\BibitemShut {NoStop}%
\bibitem [{\citenamefont {Yunker}\ \emph
  {et~al.}(2013{\natexlab{b}})\citenamefont {Yunker}, \citenamefont {Lohr},
  \citenamefont {Still}, \citenamefont {Borodin}, \citenamefont {Durian},\ and\
  \citenamefont {Yodh}}]{Yunker2013}%
  \BibitemOpen
  \bibfield  {author} {\bibinfo {author} {\bibfnamefont {P.~J.}\ \bibnamefont
  {Yunker}}, \bibinfo {author} {\bibfnamefont {M.~A.}\ \bibnamefont {Lohr}},
  \bibinfo {author} {\bibfnamefont {T.}~\bibnamefont {Still}}, \bibinfo
  {author} {\bibfnamefont {A.}~\bibnamefont {Borodin}}, \bibinfo {author}
  {\bibfnamefont {D.~J.}\ \bibnamefont {Durian}}, \ and\ \bibinfo {author}
  {\bibfnamefont {A.~G.}\ \bibnamefont {Yodh}},\ }\href@noop {} {\bibfield
  {journal} {\bibinfo  {journal} {Phys. Rev. Lett.}\ }\textbf {\bibinfo
  {volume} {110}},\ \bibinfo {pages} {035501} (\bibinfo {year}
  {2013}{\natexlab{b}})}\BibitemShut {NoStop}%
\bibitem [{\citenamefont {Kardar}\ \emph {et~al.}(1986)\citenamefont {Kardar},
  \citenamefont {Parisi},\ and\ \citenamefont {Zhang}}]{Kardar1986}%
  \BibitemOpen
  \bibfield  {author} {\bibinfo {author} {\bibfnamefont {M.}~\bibnamefont
  {Kardar}}, \bibinfo {author} {\bibfnamefont {G.}~\bibnamefont {Parisi}}, \
  and\ \bibinfo {author} {\bibfnamefont {Y.~C.}\ \bibnamefont {Zhang}},\
  }\href@noop {} {\bibfield  {journal} {\bibinfo  {journal} {Phys. Rev. Lett.}\
  }\textbf {\bibinfo {volume} {56}},\ \bibinfo {pages} {889} (\bibinfo {year}
  {1986})}\BibitemShut {NoStop}%
\bibitem [{\citenamefont {Amaral}\ \emph {et~al.}(1994)\citenamefont {Amaral},
  \citenamefont {Barabasi},\ and\ \citenamefont {Stanley}}]{Amaral1994}%
  \BibitemOpen
  \bibfield  {author} {\bibinfo {author} {\bibfnamefont {L.~A.~N.}\
  \bibnamefont {Amaral}}, \bibinfo {author} {\bibfnamefont {A.-L.}\
  \bibnamefont {Barabasi}}, \ and\ \bibinfo {author} {\bibfnamefont {H.~E.}\
  \bibnamefont {Stanley}},\ }\href@noop {} {\bibfield  {journal} {\bibinfo
  {journal} {Phys. Rev. Lett.}\ }\textbf {\bibinfo {volume} {73}},\ \bibinfo
  {pages} {62} (\bibinfo {year} {1994})}\BibitemShut {NoStop}%
\bibitem [{\citenamefont {Wakita}\ \emph {et~al.}(1997)\citenamefont {Wakita},
  \citenamefont {Itoh}, \citenamefont {Matsuyama},\ and\ \citenamefont
  {Matsushita}}]{Wakita1997}%
  \BibitemOpen
  \bibfield  {author} {\bibinfo {author} {\bibfnamefont {J.~I.}\ \bibnamefont
  {Wakita}}, \bibinfo {author} {\bibfnamefont {H.}~\bibnamefont {Itoh}},
  \bibinfo {author} {\bibfnamefont {T.}~\bibnamefont {Matsuyama}}, \ and\
  \bibinfo {author} {\bibfnamefont {M.}~\bibnamefont {Matsushita}},\
  }\href@noop {} {\bibfield  {journal} {\bibinfo  {journal} {J. Phys. Soc.
  Jpn.}\ }\textbf {\bibinfo {volume} {66}},\ \bibinfo {pages} {67} (\bibinfo
  {year} {1997})}\BibitemShut {NoStop}%
\bibitem [{\citenamefont {Myllys}\ \emph {et~al.}(2001)\citenamefont {Myllys},
  \citenamefont {Maunuksela}, \citenamefont {Alava}, \citenamefont
  {Ala-Nissila}, \citenamefont {Merikoski},\ and\ \citenamefont
  {Timonen}}]{Myllys1997a}%
  \BibitemOpen
  \bibfield  {author} {\bibinfo {author} {\bibfnamefont {M.}~\bibnamefont
  {Myllys}}, \bibinfo {author} {\bibfnamefont {J.}~\bibnamefont {Maunuksela}},
  \bibinfo {author} {\bibfnamefont {M.}~\bibnamefont {Alava}}, \bibinfo
  {author} {\bibfnamefont {T.}~\bibnamefont {Ala-Nissila}}, \bibinfo {author}
  {\bibfnamefont {J.}~\bibnamefont {Merikoski}}, \ and\ \bibinfo {author}
  {\bibfnamefont {J.}~\bibnamefont {Timonen}},\ }\href@noop {} {\bibfield
  {journal} {\bibinfo  {journal} {Phys. Rev. E}\ }\textbf {\bibinfo {volume}
  {64}},\ \bibinfo {pages} {036101} (\bibinfo {year} {2001})}\BibitemShut
  {NoStop}%
\bibitem [{\citenamefont {Maunuksela}\ \emph {et~al.}(1997)\citenamefont
  {Maunuksela}, \citenamefont {Myllys}, \citenamefont {K{\"{a}}hk{\"{o}}nen},
  \citenamefont {Timonen}, \citenamefont {Provatas}, \citenamefont {Alava},\
  and\ \citenamefont {Ala-Nissila}}]{Maunuksela1997}%
  \BibitemOpen
  \bibfield  {author} {\bibinfo {author} {\bibfnamefont {J.}~\bibnamefont
  {Maunuksela}}, \bibinfo {author} {\bibfnamefont {M.}~\bibnamefont {Myllys}},
  \bibinfo {author} {\bibfnamefont {O.-P.}\ \bibnamefont
  {K{\"{a}}hk{\"{o}}nen}}, \bibinfo {author} {\bibfnamefont {J.}~\bibnamefont
  {Timonen}}, \bibinfo {author} {\bibfnamefont {N.}~\bibnamefont {Provatas}},
  \bibinfo {author} {\bibfnamefont {M.~J.}\ \bibnamefont {Alava}}, \ and\
  \bibinfo {author} {\bibfnamefont {T.}~\bibnamefont {Ala-Nissila}},\
  }\href@noop {} {\bibfield  {journal} {\bibinfo  {journal} {Phys. Rev. Lett.}\
  }\textbf {\bibinfo {volume} {79}},\ \bibinfo {pages} {1515} (\bibinfo {year}
  {1997})}\BibitemShut {NoStop}%
\bibitem [{\citenamefont {Takeuchi}\ \emph {et~al.}(2011)\citenamefont
  {Takeuchi}, \citenamefont {Sano}, \citenamefont {Sasamoto},\ and\
  \citenamefont {Spohn}}]{Takeuchi2011}%
  \BibitemOpen
  \bibfield  {author} {\bibinfo {author} {\bibfnamefont {K.~A.}\ \bibnamefont
  {Takeuchi}}, \bibinfo {author} {\bibfnamefont {M.}~\bibnamefont {Sano}},
  \bibinfo {author} {\bibfnamefont {T.}~\bibnamefont {Sasamoto}}, \ and\
  \bibinfo {author} {\bibfnamefont {H.}~\bibnamefont {Spohn}},\ }\href@noop {}
  {\bibfield  {journal} {\bibinfo  {journal} {Sci. Rep.}\ }\textbf {\bibinfo
  {volume} {1}},\ \bibinfo {pages} {34} (\bibinfo {year} {2011})}\BibitemShut
  {NoStop}%
\bibitem [{\citenamefont {Takeuchi}\ and\ \citenamefont
  {Sano}(2010)}]{Takeuchi2010}%
  \BibitemOpen
  \bibfield  {author} {\bibinfo {author} {\bibfnamefont {K.~A.}\ \bibnamefont
  {Takeuchi}}\ and\ \bibinfo {author} {\bibfnamefont {M.}~\bibnamefont
  {Sano}},\ }\href@noop {} {\bibfield  {journal} {\bibinfo  {journal} {Phys.
  Rev. Lett.}\ }\textbf {\bibinfo {volume} {104}},\ \bibinfo {pages} {230601}
  (\bibinfo {year} {2010})}\BibitemShut {NoStop}%
\bibitem [{\citenamefont {{De Nardis}}\ \emph {et~al.}(2017)\citenamefont {{De
  Nardis}}, \citenamefont {{Le Doussal}},\ and\ \citenamefont
  {Takeuchi}}]{DeNardis2017}%
  \BibitemOpen
  \bibfield  {author} {\bibinfo {author} {\bibfnamefont {J.}~\bibnamefont {{De
  Nardis}}}, \bibinfo {author} {\bibfnamefont {P.}~\bibnamefont {{Le
  Doussal}}}, \ and\ \bibinfo {author} {\bibfnamefont {K.~A.}\ \bibnamefont
  {Takeuchi}},\ }\href {\doibase 10.1103/PhysRevLett.118.125701} {\bibfield
  {journal} {\bibinfo  {journal} {Phys. Rev. Lett.}\ }\textbf {\bibinfo
  {volume} {118}},\ \bibinfo {pages} {125701} (\bibinfo {year} {2017})},\
  \Eprint {http://arxiv.org/abs/1611.04756} {arXiv:1611.04756} \BibitemShut
  {NoStop}%
\bibitem [{\citenamefont {Fukai}\ and\ \citenamefont
  {Takeuchi}(2017)}]{Fukai2017}%
  \BibitemOpen
  \bibfield  {author} {\bibinfo {author} {\bibfnamefont {Y.~T.}\ \bibnamefont
  {Fukai}}\ and\ \bibinfo {author} {\bibfnamefont {K.~A.}\ \bibnamefont
  {Takeuchi}},\ }\href@noop {} {\bibfield  {journal} {\bibinfo  {journal}
  {Phys. Rev. Lett.}\ }\textbf {\bibinfo {volume} {119}},\ \bibinfo {pages}
  {030602} (\bibinfo {year} {2017})}\BibitemShut {NoStop}%
\bibitem [{\citenamefont {Nicoli}\ \emph {et~al.}(2013)\citenamefont {Nicoli},
  \citenamefont {Cuerno},\ and\ \citenamefont {Castro}}]{Nicoli2013}%
  \BibitemOpen
  \bibfield  {author} {\bibinfo {author} {\bibfnamefont {M.}~\bibnamefont
  {Nicoli}}, \bibinfo {author} {\bibfnamefont {R.}~\bibnamefont {Cuerno}}, \
  and\ \bibinfo {author} {\bibfnamefont {M.}~\bibnamefont {Castro}},\
  }\href@noop {} {\bibfield  {journal} {\bibinfo  {journal} {Phys. Rev. Lett.}\
  }\textbf {\bibinfo {volume} {111}},\ \bibinfo {pages} {209601} (\bibinfo
  {year} {2013})}\BibitemShut {NoStop}%
\bibitem [{\citenamefont {Yunker}\ \emph
  {et~al.}(2013{\natexlab{c}})\citenamefont {Yunker}, \citenamefont {Lohr},
  \citenamefont {Still}, \citenamefont {Borodin}, \citenamefont {Durian},\ and\
  \citenamefont {Yodh}}]{Yunker2013c}%
  \BibitemOpen
  \bibfield  {author} {\bibinfo {author} {\bibfnamefont {P.~J.}\ \bibnamefont
  {Yunker}}, \bibinfo {author} {\bibfnamefont {M.~A.}\ \bibnamefont {Lohr}},
  \bibinfo {author} {\bibfnamefont {T.}~\bibnamefont {Still}}, \bibinfo
  {author} {\bibfnamefont {A.}~\bibnamefont {Borodin}}, \bibinfo {author}
  {\bibfnamefont {D.~J.}\ \bibnamefont {Durian}}, \ and\ \bibinfo {author}
  {\bibfnamefont {A.~G.}\ \bibnamefont {Yodh}},\ }\href@noop {} {\bibfield
  {journal} {\bibinfo  {journal} {Phys. Rev. Lett.}\ }\textbf {\bibinfo
  {volume} {111}},\ \bibinfo {pages} {209602} (\bibinfo {year}
  {2013}{\natexlab{c}})}\BibitemShut {NoStop}%
\bibitem [{\citenamefont {Oliveira}\ and\ \citenamefont {{Aar{\~{a}}o
  Reis}}(2014)}]{Oliveira2014}%
  \BibitemOpen
  \bibfield  {author} {\bibinfo {author} {\bibfnamefont {T.~J.}\ \bibnamefont
  {Oliveira}}\ and\ \bibinfo {author} {\bibfnamefont {F.~D.~A.}\ \bibnamefont
  {{Aar{\~{a}}o Reis}}},\ }\href@noop {} {\bibfield  {journal} {\bibinfo
  {journal} {J. Stat. Mech.}\ }\textbf {\bibinfo {volume} {2014}},\ \bibinfo
  {pages} {P09006} (\bibinfo {year} {2014})}\BibitemShut {NoStop}%
\bibitem [{\citenamefont {Botto}\ \emph {et~al.}(2012)\citenamefont {Botto},
  \citenamefont {Yao}, \citenamefont {Leheny},\ and\ \citenamefont
  {Stebe}}]{Botto2012}%
  \BibitemOpen
  \bibfield  {author} {\bibinfo {author} {\bibfnamefont {L.}~\bibnamefont
  {Botto}}, \bibinfo {author} {\bibfnamefont {L.}~\bibnamefont {Yao}}, \bibinfo
  {author} {\bibfnamefont {R.~L.}\ \bibnamefont {Leheny}}, \ and\ \bibinfo
  {author} {\bibfnamefont {K.~J.}\ \bibnamefont {Stebe}},\ }\href {\doibase
  10.1039/c2sm25211b} {\bibfield  {journal} {\bibinfo  {journal} {Soft Matter}\
  }\textbf {\bibinfo {volume} {8}},\ \bibinfo {pages} {4971} (\bibinfo {year}
  {2012})}\BibitemShut {NoStop}%
\bibitem [{\citenamefont {Davies}\ \emph {et~al.}(2014)\citenamefont {Davies},
  \citenamefont {Kr{\"{u}}ger}, \citenamefont {Coveney}, \citenamefont
  {Harting},\ and\ \citenamefont {Bresme}}]{Davies2014}%
  \BibitemOpen
  \bibfield  {author} {\bibinfo {author} {\bibfnamefont {G.~B.}\ \bibnamefont
  {Davies}}, \bibinfo {author} {\bibfnamefont {T.}~\bibnamefont
  {Kr{\"{u}}ger}}, \bibinfo {author} {\bibfnamefont {P.~V.}\ \bibnamefont
  {Coveney}}, \bibinfo {author} {\bibfnamefont {J.}~\bibnamefont {Harting}}, \
  and\ \bibinfo {author} {\bibfnamefont {F.}~\bibnamefont {Bresme}},\
  }\href@noop {} {\bibfield  {journal} {\bibinfo  {journal} {Adv. Mater.}\
  }\textbf {\bibinfo {volume} {26}},\ \bibinfo {pages} {6715} (\bibinfo {year}
  {2014})}\BibitemShut {NoStop}%
\bibitem [{\citenamefont {Loudet}\ \emph {et~al.}(2005)\citenamefont {Loudet},
  \citenamefont {Alsayed}, \citenamefont {Zhang},\ and\ \citenamefont
  {Yodh}}]{Loudet2005}%
  \BibitemOpen
  \bibfield  {author} {\bibinfo {author} {\bibfnamefont {J.~C.}\ \bibnamefont
  {Loudet}}, \bibinfo {author} {\bibfnamefont {A.~M.}\ \bibnamefont {Alsayed}},
  \bibinfo {author} {\bibfnamefont {J.}~\bibnamefont {Zhang}}, \ and\ \bibinfo
  {author} {\bibfnamefont {A.~G.}\ \bibnamefont {Yodh}},\ }\href@noop {}
  {\bibfield  {journal} {\bibinfo  {journal} {Phys. Rev. Lett.}\ }\textbf
  {\bibinfo {volume} {94}},\ \bibinfo {pages} {018301} (\bibinfo {year}
  {2005})}\BibitemShut {NoStop}%
\bibitem [{\citenamefont {Lehle}\ \emph {et~al.}(2008)\citenamefont {Lehle},
  \citenamefont {Noruzifar},\ and\ \citenamefont {Oettel}}]{Lehle2008}%
  \BibitemOpen
  \bibfield  {author} {\bibinfo {author} {\bibfnamefont {H.}~\bibnamefont
  {Lehle}}, \bibinfo {author} {\bibfnamefont {E.}~\bibnamefont {Noruzifar}}, \
  and\ \bibinfo {author} {\bibfnamefont {M.}~\bibnamefont {Oettel}},\
  }\href@noop {} {\bibfield  {journal} {\bibinfo  {journal} {Euro. Phys. J. E}\
  }\textbf {\bibinfo {volume} {26}},\ \bibinfo {pages} {151} (\bibinfo {year}
  {2008})}\BibitemShut {NoStop}%
\bibitem [{\citenamefont {Perram}\ and\ \citenamefont
  {Wertheim}(1985)}]{Perram1985}%
  \BibitemOpen
  \bibfield  {author} {\bibinfo {author} {\bibfnamefont {J.~W.}\ \bibnamefont
  {Perram}}\ and\ \bibinfo {author} {\bibfnamefont {M.~S.}\ \bibnamefont
  {Wertheim}},\ }\href@noop {} {\bibfield  {journal} {\bibinfo  {journal} {J.
  Comp. Phys.}\ }\textbf {\bibinfo {volume} {58}},\ \bibinfo {pages} {409}
  (\bibinfo {year} {1985})}\BibitemShut {NoStop}%
\bibitem [{\citenamefont {Ara{\'{u}}jo}\ \emph
  {et~al.}(2015{\natexlab{a}})\citenamefont {Ara{\'{u}}jo}, \citenamefont
  {Schrenk}, \citenamefont {Herrmann},\ and\ \citenamefont {{Andrade
  Jr}}}]{Araujo2015b}%
  \BibitemOpen
  \bibfield  {author} {\bibinfo {author} {\bibfnamefont {N.~A.~M.}\
  \bibnamefont {Ara{\'{u}}jo}}, \bibinfo {author} {\bibfnamefont {K.~J.}\
  \bibnamefont {Schrenk}}, \bibinfo {author} {\bibfnamefont {H.~J.}\
  \bibnamefont {Herrmann}}, \ and\ \bibinfo {author} {\bibfnamefont {J.~S.}\
  \bibnamefont {{Andrade Jr}}},\ }\href@noop {} {\bibfield  {journal} {\bibinfo
   {journal} {Front. Phys.}\ }\textbf {\bibinfo {volume} {3}},\ \bibinfo
  {pages} {00005} (\bibinfo {year} {2015}{\natexlab{a}})}\BibitemShut {NoStop}%
\bibitem [{\citenamefont {Dias}\ \emph {et~al.}(2014)\citenamefont {Dias},
  \citenamefont {Ara{\'{u}}jo},\ and\ \citenamefont {{Telo da
  Gama}}}]{Dias2014a}%
  \BibitemOpen
  \bibfield  {author} {\bibinfo {author} {\bibfnamefont {C.~S.}\ \bibnamefont
  {Dias}}, \bibinfo {author} {\bibfnamefont {N.~A.~M.}\ \bibnamefont
  {Ara{\'{u}}jo}}, \ and\ \bibinfo {author} {\bibfnamefont {M.~M.}\
  \bibnamefont {{Telo da Gama}}},\ }\href@noop {} {\bibfield  {journal}
  {\bibinfo  {journal} {Europhys. Lett.}\ }\textbf {\bibinfo {volume} {107}},\
  \bibinfo {pages} {56002} (\bibinfo {year} {2014})}\BibitemShut {NoStop}%
\bibitem [{\citenamefont {Ara{\'{u}}jo}\ \emph
  {et~al.}(2015{\natexlab{b}})\citenamefont {Ara{\'{u}}jo}, \citenamefont
  {Dias},\ and\ \citenamefont {{Telo da Gama}}}]{Araujo2015}%
  \BibitemOpen
  \bibfield  {author} {\bibinfo {author} {\bibfnamefont {N.~A.~M.}\
  \bibnamefont {Ara{\'{u}}jo}}, \bibinfo {author} {\bibfnamefont {C.~S.}\
  \bibnamefont {Dias}}, \ and\ \bibinfo {author} {\bibfnamefont {M.~M.}\
  \bibnamefont {{Telo da Gama}}},\ }\href@noop {} {\bibfield  {journal}
  {\bibinfo  {journal} {J. Phys.: Condens. Matter}\ }\textbf {\bibinfo {volume}
  {27}},\ \bibinfo {pages} {194123} (\bibinfo {year}
  {2015}{\natexlab{b}})}\BibitemShut {NoStop}%
\bibitem [{\citenamefont {Zanini}\ \emph {et~al.}(2017)\citenamefont {Zanini},
  \citenamefont {Marschelke}, \citenamefont {Anachkov}, \citenamefont {Marini},
  \citenamefont {Synytska},\ and\ \citenamefont {Isa}}]{Zanini2017}%
  \BibitemOpen
  \bibfield  {author} {\bibinfo {author} {\bibfnamefont {M.}~\bibnamefont
  {Zanini}}, \bibinfo {author} {\bibfnamefont {C.}~\bibnamefont {Marschelke}},
  \bibinfo {author} {\bibfnamefont {S.~E.}\ \bibnamefont {Anachkov}}, \bibinfo
  {author} {\bibfnamefont {E.}~\bibnamefont {Marini}}, \bibinfo {author}
  {\bibfnamefont {A.}~\bibnamefont {Synytska}}, \ and\ \bibinfo {author}
  {\bibfnamefont {L.}~\bibnamefont {Isa}},\ }\href@noop {} {\bibfield
  {journal} {\bibinfo  {journal} {Nat. Comm.}\ }\textbf {\bibinfo {volume}
  {8}},\ \bibinfo {pages} {15701} (\bibinfo {year} {2017})}\BibitemShut
  {NoStop}%
\end{thebibliography}%

\end{document}